\documentclass{article}
\usepackage{graphicx}
\usepackage{amsmath}
\usepackage{amsfonts}
\usepackage{amssymb}

\begin{document}

\title{''Young'' soap films}
\author{P. G. DE GENNES\\\textit{Coll\`{e}ge de France, 11 place M.\ Berthelot}\\\textit{75231 Paris Cedex 05, France}}
\maketitle
\begin{abstract}
If we pull out rapidly a metallic frame out of a surfactant solution, we
arrive at a ''young'' soap film with relatively simple features, as noticed
first by Lucassen.\ The weight of the film is equilibrated by a vertical
gradient of surface tension.\ At each level, the local solution concentration
$c(z)$ equilibrates with the local monolayers, of surface concentration
$\Gamma(z)$.

A detailed analysis of the young films was started by us in 1987. We present
here an approach which is more illuminating \ \ \ \ a) the concentration
profiles decay exponentially at large heights, with a characteristic length
$\lambda\sim$ meters \ \ \ \ b) the surface is protected up to a thickness
$h_{m}$ larger than $\lambda$ \ \ \ \ c) we also review the dynamic
requirements.\ The surfactant must reach the surface in a time shorter than
the free fall time of a pure water film.

This discussion explains (to some extent) the compromise which is achieved in
practice by good foaming agents.
\end{abstract}

\section{Introduction}

Can a given surfactant produce a strong foam? there are some qualitative
rules. We know, for instance, that a strongly insoluble surfactant cannot
cover the surface of a rapidly growing bubble, while a soluble surfactant may
succeed -by diffusion from the bulk solution.\ We would like to make this more
precise- to characterise a surfactant solution by a few control parameters,
which tell us what are its activities.

One of the methods for generating soap films is provided by a mechanical egg
beater.\ Or, more scientifically, by pulling rapidly a metallic frame out of a
surfactant solution (fig.\label{abc}\ ) as described in the book by Mysels et
al \cite{mysels59}.%

\begin{figure}
[h]
\begin{center}
\includegraphics[
height=1.932in,
width=3.6616in
]%
{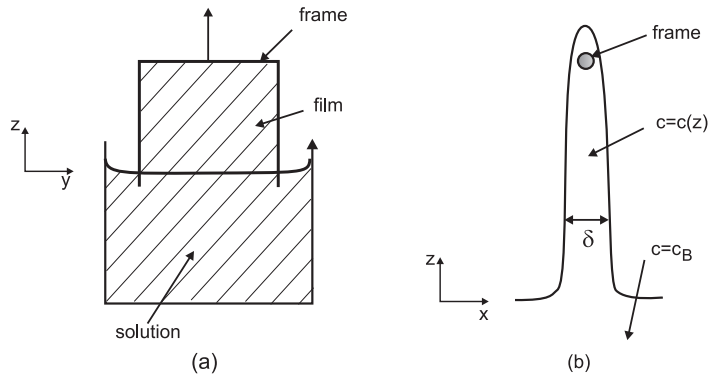}%
\caption{A ''young'' film pulled out from a surfactant solution \ \ \ \ (a)
general set up \ \ \ \ (b) detailed cross section of the film.}%
\end{center}
\end{figure}

We might first think of pulling the film at constant vertical speed.\ This is
a version of a classical Landau Levich problem \cite{landau}. But the quality
of the surfactant plays only a minor role in this process \cite{quere}: the
surfactant is pulled up at (essentially) the same speed than the solution, and
no fresh surface is generated.

A more relevant situation was invented by Lucassen \cite{lucassen81}.\ Here,
the film is pulled abruptly, and its height $h$ is comparable to (or larger
than) the horizontal span of the reservoir. Thus we must create fresh surface
by pulling the frame, and migration of surfactant from the water phase to the
surface is dominant.

This situation corresponds to what Lucassen \cite{lucassen81} called a ''young
film''. We discussed some aspects of the young films in a note \cite{pgg87}%
.\ But we missed some important points: \ \ \ \ a) we chose as our central
variable the local surface tension.\ It turns out that it is more illuminating
to study first the concentration $c(z)$ at all levels in the film \ \ \ \ b)
we did not discuss the dynamical features which tell us if the young film can
indeed be made -or not \ \ \ \ c) some serious misprints occurred in
\cite{pgg87}.

Here, we return to the static structure of the young films: in section 2 we
deal with concentrations $c$ below the vertical micelle concentration
$(c^{\ast})$.\ We find two characteristic heights: both are relevant for a
discussion of film stability.\ In section 3 we extend this to $c>c^{\ast}%
.\;$In section 4 we consider the major time constants involved: the young film
must achieve its protective surface before falling.

\section{Profile of a young film $(c<c^{\ast})$}

\subsection{Basic equations}

The film is drawn vertically (along $z$) and has a certain width $\delta(z)$
(fig. 1b).\ We assume that, in each interval $dz,$ there is a rapid
equilibration between solution (concentration $c(z)$) and surface (with
monolayer concentration $\Gamma(z)$). The total number of surfactants was
originally $c_{B}\delta$ (where $c_{B}$ is the bulk concentration in the
reservoir), and remains the same.\ Thus we must have:%

\begin{equation}
c_{B}\delta=\delta c+2\Gamma\label{eq1}%
\end{equation}

We assume that equilibrium has been achieved, and this may be written in the
differential form:%

\begin{equation}
\Gamma^{-1}d\Pi=d\mu(c)\label{eq2}%
\end{equation}

where $\Pi=\gamma_{0}-\gamma$ is the Langmuir pressure, describing the shift
of the surface tension from $\gamma_{0}$ (for pure water) to $\gamma$ (at
concentration $c$) and $\mu(c)$ is the chemical potential of the surfactant in solution.

These equations must be supplemented \ by a requirement of mechanical
equilibrium: the weight of the film must be balanced by Marangoni forces:%

\begin{equation}
\frac{-2d\Pi}{dz}=\rho g\delta\label{eq3}%
\end{equation}

($\rho=$ water density, $g=$ gravitational acceleration).

\subsection{Concentration profile}

Combining eqs (\ref{eq1}, \ref{eq2}, \ref{eq3}) we arrive at:%

\begin{equation}
\frac{-2d\Pi}{dz}\equiv-2\Pi\frac{d\mu}{dz}\frac{dc}{dz}=\frac{\rho g2\Gamma
}{c_{B}-c}\label{eq4}%
\end{equation}

In this section, we focus our attention on the dilute case $c<c^{\ast}$. Then:%

\begin{equation}
\mu=kT\;\ell nc+\text{cons}\tan\text{t}\label{eq5}%
\end{equation}

and eq. \ref{eq4} becomes:%

\begin{equation}
-\frac{dz}{\lambda}=\frac{dc}{c_{B}}\left(  \frac{c_{B}-c}{c}\right)
\end{equation}

where we have introduced a characteristic length $\lambda$ such that:%

\begin{equation}
\rho g\lambda=c_{B}\;kT
\end{equation}

$\lambda$ measures the osmotic pressure of the surfactant in terms of
hydrostatic heights.\ Note that our definition of $\lambda$ differs from ref.
\cite{pgg87}.\ Eq. (6) integrates to:%

\begin{equation}
\frac{z}{\lambda}=\frac{c-c_{B}}{c_{B}}+\ell n\left(  \frac{c_{B}}{c}\right)
\end{equation}

This concentration profile is shown on fig. 2a.\ Two interesting limits are:%

\begin{equation}
\frac{c_{B}-c}{c_{B}}=\left(  \frac{2z}{\lambda}\right)  ^{1/2}\qquad
(z<\lambda)
\end{equation}%

\begin{equation}
\frac{c}{c_{B}}=e^{-z/\lambda}\qquad(z>\lambda)
\end{equation}%

\begin{figure}
[h]
\begin{center}
\includegraphics[
height=2.0245in,
width=3.0502in
]%
{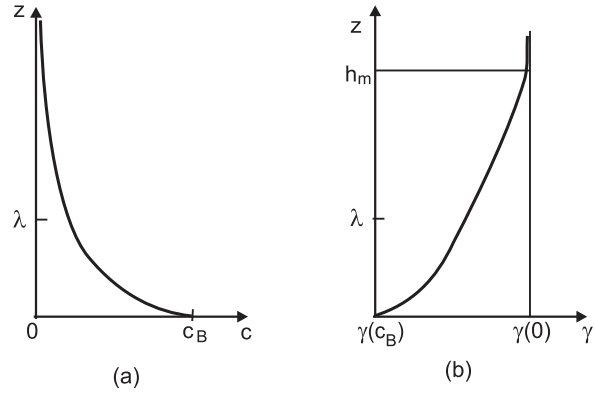}%
\caption{Structure of a young film when the bulk concentration ($c_{B}$) is
smaller than the $cmc(c^{\ast})$ \ \ \ \ (a) concentration profile \ \ \ \ (b)
variation of surface tension with height.}%
\end{center}
\end{figure}

\subsection{Thickness profile}

We can translate $c(z)$ into a thickness profile $\delta(z)$ using eq. (1). In
most of the region $c<c^{\ast}$, the monolayer concentration $\Gamma$ is
nearly constant (adsorption has taken place abruptly at very low $c^{\prime}%
s$). Then we have:%

\begin{equation}
\delta=\frac{2\Gamma_{B}}{c_{B}-c}%
\end{equation}

where $\Gamma_{B}$ is $\Gamma(c=c_{B}).$ The thickness $\delta(z)$ described
by eq. (11) is a decreasing function of $c$. When $z>\lambda$, we have
$c<<c_{B},$ and $\delta$ reaches a simple limit $\delta=2\ell$, where:%

\begin{equation}
\ell=\Gamma_{B}/c_{B}%
\end{equation}

The length $\ell$ plays an important role in foaming problems: $\ell$ is the
minimal thickness of solution required to transfer surfactant from bulk to
surface, achieving the required surface concentration $\Gamma_{B}.$

\subsection{Vertical variations of the surface tension $\gamma$}

We know $c(z)$, and we know (from classical plots for usual surfactants) the
surface tension $\gamma(c)$: thus we may construct $\gamma(z)$=$\gamma_{0}%
-\Pi(z).$ The general aspect of this curve is shown on fig.\ 2b.

For $z>\lambda$ ($c<<c_{B}$), we may write:%

\begin{equation}
\frac{d\gamma}{dz}=-\frac{d\Pi}{dz}=\frac{\rho g\Gamma}{c_{B}-c}\cong
\frac{\rho g\Gamma_{B}}{c_{B}}=\text{constant}%
\end{equation}

Thus $\gamma(z)$ increases linearly with $z$, up to a certain height ($h_{m}
$), where we return to a bare surface ($\Gamma\sim0,$ $\gamma=\gamma_{0}%
$).\ This corresponds to:%

\[
h_{m}\frac{d\gamma}{dz}\sim\gamma_{0}-\gamma_{B}\equiv\Pi_{B}%
\]

or:%

\begin{equation}
h_{m}=\lambda\frac{\Pi_{B}}{kT\;\Gamma_{B}}%
\end{equation}

Note that $h_{m}$ is larger than $\lambda$,because $\Pi_{B}$ si much larger
than an ideal gas pressure. The height $h_{m}$ is an absolute limit of
stability for our young film. (But of course, other instabilities may occur in
the regime of high $z$, low $\delta$).

\section{Extension to higher concentrations $c_{B}>c^{\ast}$}

\subsection{Structure of the chemical potential at $c>>c^{\ast}$}

At concentrations $c>c^{\ast}$, most of the surfactant is in a micellar form
(concentration $c_{M}$).\ Only a small fraction ($c_{1}$) is present in a
monomer form. Let us call $N$ the number of surfactants per micelle.\ Then a
rough but convenient description (with $N$ fixed) of the micelle/monomer
equilibrium, may be written in the form:%

\begin{equation}
c_{M}=\frac{c_{1}^{N}}{c^{\ast N-1}}%
\end{equation}

or for the total concentration:%

\begin{equation}
c=c_{1}+c_{M}=c_{1}\left\{  1+\left(  \frac{c_{1}}{c^{\ast}}\right)
^{N-1}\right\}
\end{equation}

The chemical potential is still:%

\begin{equation}
\mu=kT\;\ell n\;c_{1}+\text{cons}\tan\text{t}%
\end{equation}

Let us consider only the limit $c>>c^{1}$.\ Then:%

\begin{equation}
\frac{c}{c^{1}}\sim\frac{c_{M}}{c^{\ast}}=\left(  \frac{c_{1}}{c^{\ast}%
}\right)  ^{N}%
\end{equation}

and:%

\begin{equation}
\mu=kT\left\{  \ell n\;c^{\ast}+\frac{1}{N}\ell n\frac{c}{c^{\ast}}\right\}
\end{equation}%

\begin{equation}
\frac{d\mu}{dc}=\frac{kT}{Nc}%
\end{equation}

\subsection{Concentration profile}

Returning to eq. (4), we see from eq. (20) that for $c>>c^{\ast},$ we should
replace $\rho g$ by $N\rho g.$ Or, equivalently, the characteristic length in
the high concentration regime is:%

\begin{equation}
\overset{\sim}{\lambda}=\frac{\lambda}{N}%
\end{equation}

This is much smaller than $\lambda$ since $N$ is large ($\sim$ 60).

The resulting plot of $c(z)$ is shown on fig.\ 3a. We start at $z=0$ from a
concentration $c_{B}>>c^{\ast}.\;$Then upon increasing $z$, the concentration
decreases, and when $c<<c_{B}$, we have simply:%

\begin{equation}
c(z)=c_{B}\exp\left(  -\frac{Nz}{\lambda}\right)  \qquad(z<z^{\ast})
\end{equation}%

\begin{figure}
[ptb]
\begin{center}
\includegraphics[
height=1.9839in,
width=2.9879in
]%
{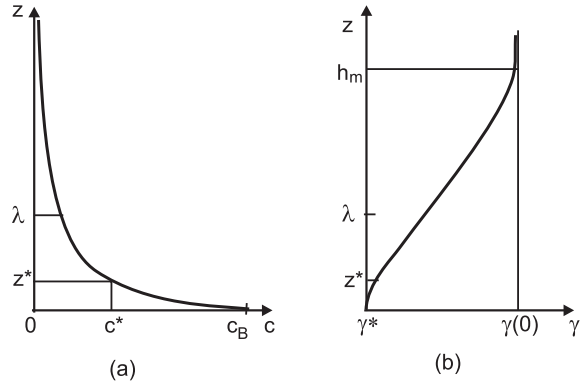}%
\caption{Structure of a young film when $c_{B}>>c^{\ast}$ \ \ \ \ (a)
concentrations \ \ \ \ (b) surface tensions}%
\end{center}
\end{figure}

This holds up to a height $z^{\ast}$ where $c=c^{\ast}:$%

\begin{equation}
z^{\ast}=\frac{\lambda}{N}\ell n\frac{c_{B}}{c^{\ast}}%
\end{equation}

At $z>z^{\ast}$, we return to the low concentration regime (eq. 6), and
ultimately, for $z>\lambda$, we have:%

\begin{equation}
c(z)\sim c^{\ast}\exp\left(  -\frac{z}{\lambda}\right)
\end{equation}

\subsection{Surface tensions}

In all the region $c>c^{\ast},$ or $z<z^{\ast}$, the surface tension keeps its
minimum value $\gamma^{\ast.}.\;$Above $z<z^{\ast},$ we return to the
cons$\tan$t slope regime of eq. (13).\ The maximum height allowable $h_{m}$ is
still given by eq.\ (14) as shown on fig. (3b).

Note the inequalities:%

\begin{equation}
h_{m}>\lambda>z^{\ast}%
\end{equation}

\section{Discussion}

\subsection{Summary of static predictions:}

\qquad a) in a young film, we expect a concentration $c(z)$ which decays
exponentially at large $z$, with a characteristic length $\lambda$ (eq.\ 7).
$\lambda$ is typically a few meters.\ This statement holds for bulk
concentrations $c_{B}$ which may be lower or higher than $c^{\ast}$.

\qquad b) the film thickness at high $z$ is simply $\delta=2\ell$, where
$\ell=\Gamma_{B}/c_{B}$ \ \ \ \ -$(c<c^{\ast})$ or $\ell=\Gamma_{B}/c^{\ast}$
for $c>c^{\ast}.$

\qquad c) the film is bare and completely unstable at heights $z>h_{m}$, where
$h_{m}$ is given by eq.\ (14), and is larger than $\lambda.$

\qquad d) on the whole, going to bulk concentrations, $c_{B}>c^{\ast}$ does
not produce great alterations: most static film properties should be close to
what they are for $c_{B}=c^{\ast}.$

\subsection{Characteristic times}

1) A water film with no surfactant will collapse under its own weight in a
time $t_{c}$ governed by gravitational\ \ flow with the acceleration $g:$%

\begin{equation}
t_{c}\sim\sqrt{\frac{h}{g}}%
\end{equation}

where $h$ is the height of the film.\ For $h=10$ cm, \ \ \ \ $t_{c}\sim0.1$ sec.

2) What is the time required for the surfactant to reach the surface? We
consider here first the (relatively simple) case where $c_{B}<c^{\ast}$: all
the transport is via monomers.\ We also assume that there is no barrier at the
surface opposing the surfactant adsorption.\ Then, we must transfer the
surfactant by diffusion (coefficient $D$) over a distance $\ell$ defined in
eq.\ (12).\ The diffusion time is roughly:%

\begin{equation}
t_{d}=\frac{\ell^{2}}{D}=\frac{\Gamma_{B}^{2}}{c_{B}^{2}D}%
\end{equation}

The condition of formation of a young film at $c<c^{\ast}$ is essentially that
$t_{d}$ be shorter than $tc$. To get small $t_{d}^{\prime}s$, we must use
concentrations $c_{B}$ which are as high as possible (but remain below
$c^{\ast}$).

3) The extension of these ideas to $c_{B}>c^{\ast}$ is quite delicate.
Micelles provide a big reservoir of surfactant, but the delivery is slow.

\qquad a) micelles can emit monomers which are then absorbed by the fresh
surface: this process is fast (10$^{-6}$ seconds) but inefficient, because
each micelle cannot give more than 2 or 3 monomers (beyond that, the micellar
free energy rises fast).

\qquad b) micelles can split in two, and the corresponding characteristic time
$\tau_{split}$ is relatively long (10$^{-3}$ sec).\ After splitting, the parts
end up as monomers, and can feed the surface. This process has been documented
by Shah \cite{shah}.

\qquad c) the micelles may unfold by direct contact with the surface. The
energy barriers for processes (b) and (c) are comparable -maybe the barrier
for (c) is slightly lower.

If, for simplicity, we focus on process (b), the conclusion is that if we have
$t_{split}<t_{c}$, the young film should be realisable.\ The opposite limit
($t_{split}>t_{c}$) would occur only for very long surfactant chains.

\subsection{Conclusions}

\qquad a) To achieve young films of height $h$, we need $h_{m}>h$, where
$h_{m}$ is given by eq. (14).\ As shown by this equation, high $h_{m}$
corresponds to high surface pressures (or $\gamma<<\gamma_{0}$) in the present
solution.\ To achieve this, our surfactants must have an aliphatic tail which
is not too short.

\qquad b) The surfactant must migrate fast to the water/air interface. For
$c\sim c^{\ast}$, this implies that $\ell^{\ast}=\Gamma^{\ast}/c^{\ast}$ must
be relatively small.\ The critical micelle concentration $c^{\ast}$ should not
be too small.\ The surfactant tails should not be too long.

Thus a good foaming agent results from a compromise: this does correspond to
the empirical rule stating that the HLB\ should be in the range 7 to
9.\bigskip\bigskip

{\large Acknowledgments}: we have benefited from useful advice by D.\ Langevin
and M.\ Schott.

\end{document}